\documentclass[twocolumn,prb,amssymb,showpacs,showkeys]{revtex4}
\usepackage{graphicx}
\usepackage{dcolumn}
\usepackage{bm}
\begin{document}

\title{Pseudogap behavior in Pr$_{0.5}$Sr$_{0.5}$MnO$_3$: A photoemission study}
\author{P. Pal}
\email{prabir.pal@maxlab.lu.se}
\affiliation{Department of Physics and Astronomy, Uppsala University, Box 516, SE 751 21 Uppsala, Sweden.}
\affiliation{MAX-lab, Lund University, Box 118, SE-221 00, Lund, Sweden.}

\author{M. K. Dalai}
\affiliation{National Physical Laboratory, Dr. K. S. Krishnan Marg, New Delhi-110 012, India.}

\author{I. Ulfat}
\affiliation{MAX-lab, Lund University, Box 118, SE-221 00, Lund, Sweden.}
\affiliation{Department of Applied Physics, Chalmers University of Technology, SE-412 96, G\"oteborg, Sweden.}

\begin{abstract}

The valence band electronic structure of Pr$_{0.5}$Sr$_{0.5}$MnO$_3$ has been investigated across its paramagnetic metallic (PMM) - ferromagnetic metallic (FMM) - antiferromagnetic insulator (AFMI) transition. Using surface sensitive high resolution photoemission we have conclusively demonstrated the presence of a pseudogap of magnitude $80$ meV in the near Fermi level electronic spectrum in the PMM and FMM phases and finite intensity at the Fermi level in the charge ordering (CO)-AFMI phase. The pseudogap behavior is explained in terms of the strong electron-phonon interaction and the formation of Jahn Teller (JT) polarons, indicating the charge localizations. The finite intensity at the Fermi level in the insulating phase showed a lack of charge ordering in the surface of the Pr$_{0.5}$Sr$_{0.5}$MnO$_3$ samples.

\end{abstract}

\maketitle

\section{\bf INTRODUCTION}

The perovskite manganese oxides exhibit diversity in electronic, magnetic and structural transition. A notable feature of charge ordering (CO) resulting from the dominance of on-site coulomb interaction over the kinetic energy of the charge carriers \cite{shenoy1,shenoy2} has been observed for the half-doped manganites. It is quite fascinating due to the interesting basic physics underlying the phenomena. Although, in general, it is believed that the phenomena is based on strong electron - phonon coupling that results in polaron formations \cite{millis1,millis2,aliaga} along with the double exchange (DE) interactions \cite{zener,anderson}, the detailed theoretical explanation is still under debate. In a microscopic scenario the polaron effect is considered to be intricately coupled to the CO phenomenon \cite{ahn}. The prominent role of CO in manganites originates from the occurrence of a well pronounced metal-insulator transition associated with the magnetic phase transitions.

Among the half-doped manganites, Pr$_{0.5}$Sr$_{0.5}$MnO$_3$ not only exhibits CO state across its ferromagnetic metallic (FMM) to antiferromagnetic insulating (AFMI) transition but also multiple magnetic phase transitions upon cooling resulting in alluring structural changes \cite{llobet}. Moreover the magnetic phase transitions are expected to modify the near Fermi level (E$_F$) electronic structure of these materials, especially when accompanied by a change in the lattice symmetry \cite{llobet}. This motivates the study of electronic structure of the energy scales involved in the CO phenomena and the associated changes in the near E$_F$ electronic structure.

Photoemission spectroscopy is one of the most powerful experimental techniques to extract information in the low energy spectral weight changes near the Fermi level (E$_F$). Pr$_{0.5}$Sr$_{0.5}$MnO$_3$ has been studied earlier using photoemission spectroscopy \cite{kurmaev,chainani,pal1}. However the majority of these studies are emphasized on bulk sensitive angle integrated photoemission which just determine the bulk spectral density of states (DOS), but not the surface spectral DOS. By using high resolution surface sensitive ultraviolet photoemission spectroscopy (UV-PES) we have found the formation of a pseudogap of energy scale $80$ meV in the PMM and FMM phases and a finite intensity at the Fermi level in the CO-AFMI phase. The UV-PES (h$\nu$ $=$ $50$ eV) spectra with a short electron mean free path mainly reflect the surface electronic structures, which deviate from the bulk Mn $3d$ states. In the surface, the CO state competes with the ferromagnetic fluctuations owing to which the photoemission results are quite different from those of bulk states at low temperatures. So far no measurements are available showing the formation of pseudogap in the PMM and FMM phase and a gapless excitations in the CO-AFMI phase using surface sensitive PES. Our studies, presented in this paper, show some new as well as improved results on the electron energy scales which will have substantial importance to the understanding of the lack of CO in the first few layers from the surface.

\section{\bf EXPERIMENT}

The single-crystal samples of Pr$_{0.5}$Sr$_{0.5}$MnO$_3$ were prepared by the floating zone method in a mirror furnace. The compositional homogeneity of the crystal was confirmed using energy-dispersive spectroscopy analysis. Room temperature powder X-ray diffraction pattern confirmed the tetragonal structure (space group I4/mcm). Temperature dependence of magnetization and resistivity were measured to confirm the multiple transitions. From a paramagnetic metallic (PMM) state at room temperature, this composition turns to a FMM state below T$_C$ ($\sim$ $250$ K) and finally to a CO - AFMI state at T$_N$ ($\sim$ $150$ K) \cite{kawano}. The PMM - FMM phase transition is isostructural (tetragonal with c/a $>$ $1$) while the FMM - AFMI transition leads to monoclinic change in lattice symmetry.

The high resolution photoemission (PES) measurements were performed on the Pr$_{0.5}$Sr$_{0.5}$MnO$_3$ samples at I511 beamline, MaxLab, Lund, Sweden, using a Scienta $R4000$ analyzer. Photoelectrons were collected with an angle-integrated mode, with an acceptance angle of $\pm$ 19$^\circ$ around the normal to the sample surface. The photon energy $50$ eV was used for the PES measurements. The binding energy was calibrated by using the Fermi-edge of the gold reference sample and the energy resolution is set at $20$ meV for $50$ eV photons. The base pressure of the chamber was $\sim$ $1.0$ $\times$ $10^{-10}$ mbar. Temperature of the sample was measured using a calibrated silicon diode sensor and was controlled using a local heater. To check reproducibility of temperature dependent data, we scraped the sample at room temperature and then measured in a temperature cycle from $352$ to $30$ K and back to $352$ K. Scraping was repeated until negligible intensity was found for the bump around 9.5 eV, which is a signature of surface contamination or degradation \cite{sarma}. The spectra were also obtained on several fractured samples cut from the same ingot. It is important to stress that no significant changes are observed in the valence band region cleaned by in-situ scraping or fracture. The spectra were collected within $2$ h of scraping or fracture and no changes in the spectra were observed due to surface degradation or contamination within this period.

\section{\bf RESULTS AND DISCUSSION}

\begin{figure}[t]
\includegraphics[width=3.0in]{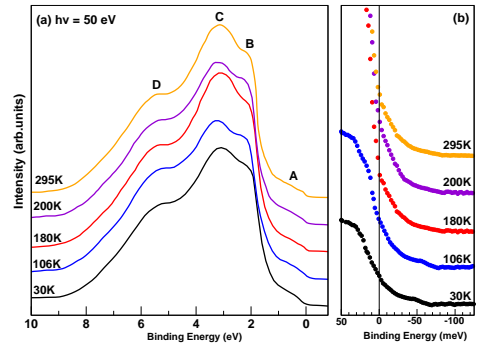} 
\caption{\label {figure_1}(Color online) (a) Valence band photoemission spectra of Pr$_{0.5}$Sr$_{0.5}$MnO$_3$ obtained using 50 eV photon energy across the PMM-FMM-AFMI phase transition. (b) Expanded spectra showing the spectral changes in the vicinity of E$_F$.}
\end{figure}

Fig. 1(a) shows the valence band photoemission spectra of the Pr$_{0.5}$Sr$_{0.5}$MnO$_3$ sample obtained by using a photon energy of $50$ eV across its PMM-FMM-AFMI transitions. The spectra exhibit four distinct features marked A ($\sim$ $1.1$ eV), B ($\sim$ $2.1$ eV), C ($\sim$ $3.1$ eV) and D ($\sim$ $5.4$ eV). Earlier experiments \cite{pal1,dalai,ebata,pal2} and band structure calculations \cite{monodeep} have shown that the feature A originates from the e$_{g}$ states and B and C are from the strongly mixed Mn 3d t$_{2g}$, O 2p, and Pr 4f states. Feature D has a dominant O $2p$ character with a small Mn $3d$ contribution. The spectra from the PMM and FMM phases show a clear Fermi edge. It is also clear from the figure that the width of the feature A (which corresponds to the e$_{g}$ states) decrease as we go down in temperature across the FMM-AFMI transition. Further, the intensity of the near E$_F$ was found to decrease as the material goes through this transition [Fig.1(b)]. This could be due to the localization of the e$_g$ electrons in the CO-AFMI phase. Further, a small but finite intensity at E$_F$ was observed in the insulating state of the sample. The electron mean free path has a minimum at about $40-50$ eV electron kinetic energy. The spectra were also obtained using $h\nu=70$ eV. We note that with $h\nu=50$ eV (below the Mn $3p$-$3d$ threshold) the intensity at Fermi level is more than that of the spectra measured at $70$ eV of photons. This confirm that the intensity at E$_F$ in the electronic structure increases in the surface and decreases systematically in the bulk. Considering the difference in the electron mean free paths between the bulk and surface sensitive photoemission spectra, our results indicate that the surface Mn $3d$ states are somewhat different from that of the bulk Mn $3d$ states in the Pr$_{0.5}$Sr$_{0.5}$MnO$_3$. Bulk photoemission studies exhibit a CO gap at low temperature \cite{kurmaev,chainani} whereas surface sensitive photoemission measurements indicate a finite photoemission intensity at E$_F$ in the CO phase. It is to be noted that in-situ scraping or fracture will remove one out-of-plane oxygen (Mn-O-Mn bond) in the MnO$_6$ octahedra. This structural change in the surface can lead to an increase in the other out-of-plane Mn $3d$-O $2p$ hybridization strength due to one short bond and a consequent increase in the one electron band width (W). Thus it is justified to believe that a ferromagneic fluctuation of itinerant metallic nature of the Mn e$_g$ band in the surface competes with the local coulomb interaction. Consequently, the CO state in the surface is suppressed from that of the bulk and the coulomb interactions were found quite ineffective in the surface at low temperatures. This scenario is further supported by the observation of significant intensity at the Fermi level. The finite intensity at the Fermi level suggests the rotation and distortion of the MnO$_6$ octahedra in the insulating phase giving rise to the changes in the crystal structure which is reflected strongly on its surface. We propose that the structure of first two layers from the surface may be different due to reduced atomic coordinations, which is responsible for the observed finite intensity at E$_F$ at the low temperature CO phase. In this regard it is important to mention that the earlier studies on related manganites using UV-PES also reported a finite intensity at E$_F$ in the insulating state \cite{wadati}.

\begin{figure}[t]
\includegraphics[width=3.0in]{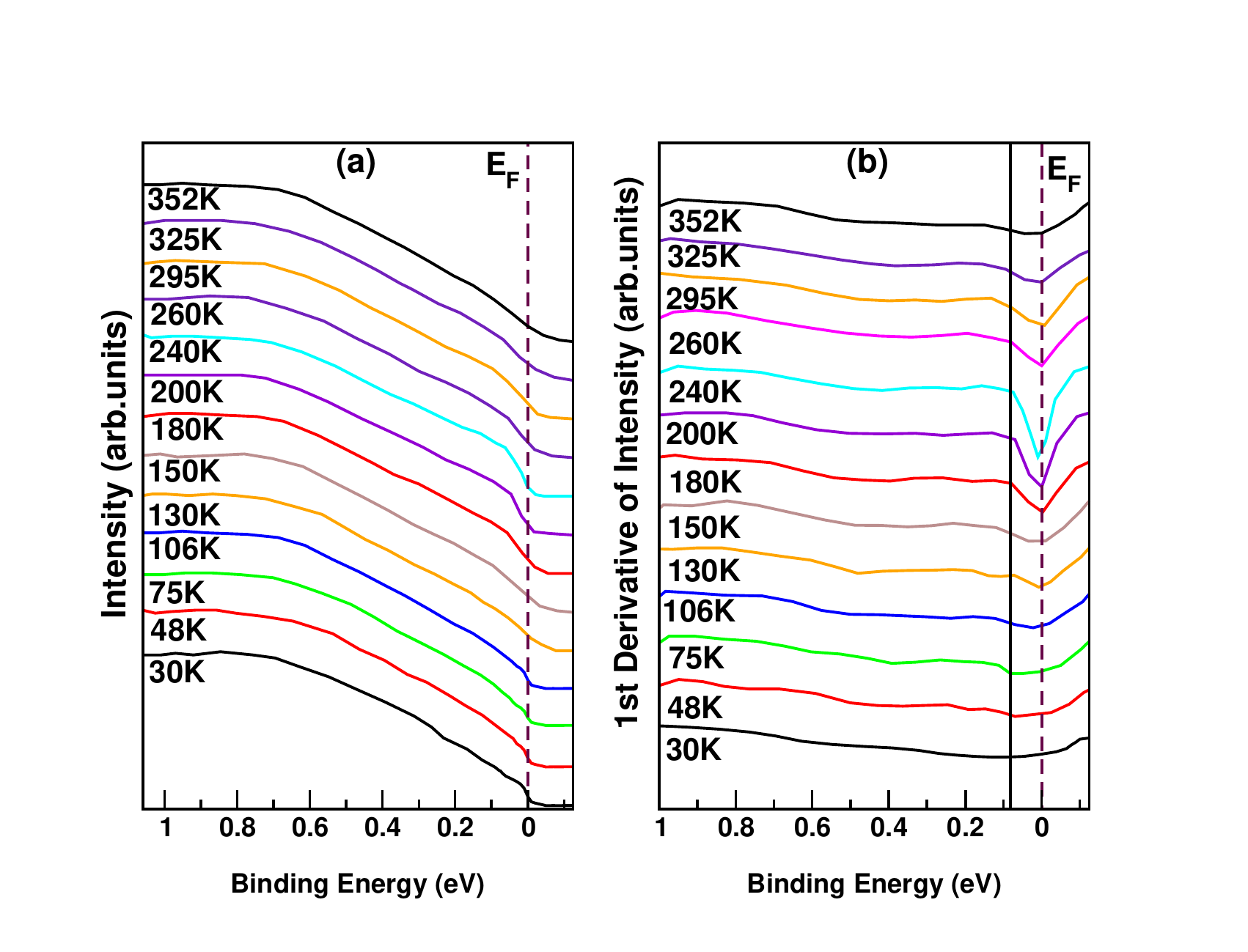} 
\caption{\label {figure_2} (Color online) (a) High-resolution valence band photoemission spectra of the near E$_F$ region obtained by using $50$ eV photons. (b) First derivative of the spectra shown the panel (a) to highlight the temperature dependent dip near E$_F$ resulting the formation of a pseudogap. (The temperature dependent dip near E$_F$ corresponds to the energy scale of $80$ meV marked by the solid line, measures the pseudogap.)}
\end{figure}

We have further analyzed the near E$_F$ spectral behavior across the PMM - FMM - AFMI transition. Panel (a) of Fig. 2 shows the high resolution near E$_F$ photoemission spectra from the Pr$_{0.5}$Sr$_{0.5}$MnO$_3$ sample taken at a photon energy 50 eV. All the spectra at different temperatures are normalized at energies above $1.0$ eV from E$_F$ and then shifted along y-axis for clarity. The finer changes in the near E$_F$ region are depicted in panel (b) where the first-order derivative of the intensity is plotted against the binding energy. The FMM-AFMI transition, as expected, shows a smearing out of the Fermi edge. However, at temperatures below $352$ K (PMM phase) a dip at the E$_F$ starts to form gradually and becomes sharper through the FMM phase till the T$_N$. This dip in the derivative spectra is a signature of a temperature dependent pseudogap existing in the PMM and FMM phases. The pseudogap is formed due to the transfer of spectral weight from the E$_F$ to higher binding energy positions with lowering of temperature. Our spectra show the energy scale over which a sharp dip in the FMM and the PMM phases occurs to be $\sim$ $80$ meV. This $80$ meV pseudogap-like feature is indicated by the solid line in the figure 2(b). Some indications of its temperature dependence are already obvious in the raw data and in the first derivative spectra. In order to confirm the pseudogap-like feature, we symmetrize the spectra and plot with a constant vertical shift as shown in Fig. 3(a). The numerical procedure removes any thermal broadening contribution arising from the Fermi function which are symmetric about the zero of the binding energy scale and gives the change in the density of states within 5K$_B$T of E$_F$.  The analysis shows an interesting evolution with temperature, which leads us to identify a pseudogaplike structure at E$_F$ of $\sim80$ meV. In Fig. 3(b), we have plotted the symmetrized spectra corresponding to each temperature together with that from the T$=200$ K for a comparison. This figure shows that there are drastic changes in the electronic structure within $\sim$ 80 meV of E$_F$ and the pseudogaplike feature is more pronounced in the FMM and PMM phases. Thus the energy scale involved in the formation of pseudogap behavior ($80$ meV) is comparable to the value of $100$ meV CO gap observed by Chainani et al. \cite{chainani}. Across the FMM-AFMI transition, a finite intensity at the Fermi level is observed as we go down in temperature which suggests that the electronic structure of first few layers from the surface is different from that of the bulk. Our high resolution spectra show the existence of pseudogap in the PMM and FMM phases which is important for the metal-insulator transition and a gapless finite intensity at the E$_F$ in the CO-AFMI phase. As mentioned earlier, the crystal structure of the Pr$_{0.5}$Sr$_{0.5}$MnO$_3$ is tetragonal in both the PMM and FMM phases. Further, both these states have an A-type orbital ordering where the Mn d$_{x^2-y^2}$ orbitals are ordered and strongly hybridized with the O $2p$ orbitals \cite{kajimoto}. Finally with decreasing temperature from the PMM phase to the FMM phase this orbital ordering seems to be strengthened along with the magnetic ordering. This strengthening of the hybridization between the in-plane Mn d$_{x^2-y^2}$ orbitals and the O $2p$ orbitals can make the MnO$_6$ octahedra elongated in the Z-axis. Such a distortion of the MnO$_6$ octahedra may shift the out-of-plane Mn 3d$_{z^2-r^2}$ orbitals where the e$_g$ electron is occupied, up in energy; like in the Q-modes of the Jahn Teller (JT) polaron \cite{millis1}. The spectral weight behavior and the consequent pseudogap in the PMM and FMM phases, found here, indicate the role of a strong electron - phonon interaction and formation of JT polarons.

\begin{figure}[t]
\includegraphics[width=3.0in]{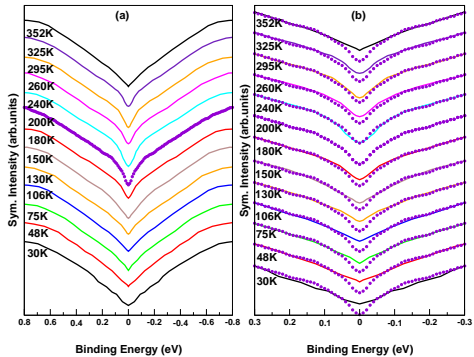} 
\caption{\label {figure_3}(Color online) (a) Symmetrized spectra shows a pseudogap at E$_F$. (b) All of the other symmetrized spectra plotted against the one from T$=$ 200K data, depicting the temperature dependent dip at E$_F$.}
\end{figure}

\section{CONCLUSION}

Our study of the temperature dependent magnetic phase transitions in Pr$_{0.5}$Sr$_{0.5}$MnO$_3$ compounds conclusively shows a pseudogap behavior in their near E$_F$ electronic spectrum over a large region of their phase diagram and a finite photoemission intensity at E$_F$ in the CO-AFMI state. The intensity at E$_F$ in the insulating phase is ascribed to the lack of CO in the first few layers from the surface. The formation of pseudogaps were discussed considering the strong electron - phonon interaction, consequent charge localizations and possible formation of JT polarons.

\section{Acknowledgments}

 P. P. is grateful to Carl Trygger Foundation for financial support.

\end{document}